\documentclass[12pt]{article}
\usepackage{graphicx,cite,amsmath,amssymb}

\usepackage{a4wide}

\def\De{\Delta}

\newcommand{\MW}{M_W}
\newcommand{\MZ}{M_Z}

\newcommand{\sweff}{\sin^2\theta_{\mathrm{eff}}}

\newcommand{\TeV}{\,\, \mathrm{TeV}}

\newcommand{\GeV}{\,\, \mathrm{GeV}}
\newcommand{\mev}{\,\, \mathrm{MeV}}
\newcommand{\MeV}{\,\, \mathrm{MeV}}
\def\de{\delta}
\def\Si{\Sigma}

\newcommand{\mt}{m_{t}}

\newcommand{\mb}{m_{b}}

\def\reffi#1{\mbox{Fig.~\ref{#1}}}

\newcommand{\gsim}
{\;\raisebox{-.3em}{$\stackrel{\displaystyle >}{\sim}$}\;}
\newcommand{\VL}{\left( \begin{array}{c}}
\newcommand{\VR}{\end{array} \right)}

\newcommand{\MLv}{\left( \begin{array}{cccc}}
\newcommand{\MR}{\end{array} \right)}
\newcommand{\ML}{\left( \begin{array}{cc}}
\newcommand{\MLd}{\left( \begin{array}{ccc}}

\newcommand{\tb}{\tan \beta}

\def\_{\rule{.3em}{.15ex}}

\def\slash#1{\setbox0=\hbox{$#1$}#1\hskip-\wd0\dimen0=5pt\advance
       \dimen0 by-\ht0\advance\dimen0 by\dp0\lower0.5\dimen0\hbox
         to\wd0{\hss\sl/\/\hss}}

\begin{document}
\thispagestyle{empty}

\def\thefootnote{\fnsymbol{footnote}}

\begin{flushright}
CERN--PH--TH/2005-153\\
UB--ECM--PF--05/20\\
hep-ph/0508241 \\
\end{flushright}

\vspace{1cm}

\begin{center}
{\large\sc {\bf $\MW$ and $\sweff$ in Split SUSY: present and future
    expectations}} 
\vspace{1cm}

{\sc 
Jaume Guasch$^{1}$%
\footnote{email: guasch@ecm.ub.es}%
and Siannah Pe\~naranda$^{2}$%
\footnote{email: siannah.penaranda@cern.ch}
}

\vspace*{1cm}

{\sl
$^1$Departament d'Estructura i Constituents de la
    Mat{\`e}ria, Facultat de F{\'\i}sica,\\ 
Universitat de Barcelona, Diagonal 647, E-08028 Barcelona, Catalonia,
Spain.

\vspace*{0.4cm}

$^2$CERN TH Division, Department of Physics,\\ 
CH-1211 Geneva 23, Switzerland
}

\end{center}

\vspace*{0.2cm}

\begin{abstract}
\noindent 
We analyse the precision electroweak observables $\MW$ and $\sweff$
and their correlations in the recently 
proposed Split SUSY model. 
We compare the results with the Standard Model and Minimal
Supersymmetric Standard Model predictions, and with present and future
experimental accuracies. Present experimental accuracies in
($\MW$, $\sweff$) do not allow constraints to be placed on the Split SUSY
parameter space. We find that the shifts in ($\MW$, $\sweff$) induced by
Split SUSY 
can be larger than the anticipated accuracy of the GigaZ option of the
International Linear Collider, and that the most sensitive observable is
$\sweff$. These large shifts are possible also for large chargino masses
in scenarios with small $\tb\simeq1$.
\end{abstract}

\noindent\textbf{PACS:} 12.15.Lk, 12.60.Jv, 14.70.Fm, 14.70.Hp

\def\thefootnote{\arabic{footnote}}
\setcounter{page}{0}
\setcounter{footnote}{0}

\newpage


The Standard Model (SM) with minimal Higgs-field content
could turn out not to be the basic 
theoretical framework for describing electroweak symmetry breaking.
During the last decades, Supersymmetry (SUSY) has become one of the most promising
theoretical ideas beyond the SM.
The Minimal Supersymmetric Standard Model
(MSSM)~\cite{susy} is the simplest supersymmetric
extension of the SM, and it is at least as successful as the SM
to describe the experimental data~\cite{hollik}.
This model predicts the existence of scalar partners $\tilde{f}_{\rm L}$, 
$\tilde{f}_{\rm R}$ to each SM chiral fermion, and of spin-1/2 partners to the 
gauge and Higgs bosons.
It is found that the effects of SUSY at the scale of ${\cal{O}}({\rm TeV})$
can provide a theoretically well motivated solution to the hierarchy
problem and also predicts the unification of the gauge
couplings~\cite{Dimopoulos,Dimopoulos-uni}. 
Moreover, the lightest neutralino in SUSY models constitutes a
promising dark matter candidate~\cite{Dimopoulos}. 
However, in spite of the above successes, SUSY still has some unsolved
problems for phenomenological reasons. For instance, large flavour
mixing and proton decay, as well as a too large  cosmological constant,
are predicted by these models. 

Recently, the scenario of Split SUSY has been 
suggested~\cite{split1,split2,split3}. In
this scenario, the SUSY-breaking scale is much heavier than the
electroweak scale, i.e. there is a hierarchy between the scalar
superpartners and the fermionic partners of the SM particles. Therefore, 
except for one Higgs-boson, all scalar particles (squarks, sleptons and 
extra MSSM Higgs particles) are very heavy, of the order
of $10^9\GeV$, while the fermions (gauginos and higgsinos) are
kept at the electroweak scale. Thus, only the SM spectrum, including
one Higgs scalar, and gauginos and higgsinos remain.
The rest of the MSSM spectrum decouples~\cite{siannah}. 
This scenario implies the existence of an ``unnatural'' 
fine-tuning, such that the Higgs-boson vacuum expectation value can be
kept at the
observed electroweak scale. Assuming this fine-tuning effect, some of
the remaining problems in SUSY models are solved: 
as a consequence of decoupling of all sfermions,
there is no flavour-changing neutral current problem that emerges in
the MSSM, and the mediating proton decay problem has been eliminated. 
On the other hand, keeping gauginos and higgsinos at the
electroweak scale, gauge unification is preserved and we can have a
neutralino as a good candidate for dark matter. 

Phenomenological implications of Split SUSY have been extensively  
discussed during the last year~\cite{Dimopoulos-susy05}. 
An alternative way, with respect to the direct search for beyond 
the SM physics or Higgs
particles, is to probe new physics through virtual effects of the 
additional non-standard particles to precision observables. 
In particular, the analysis of radiative effects of light gauginos and
higgsinos to precision electroweak (EW) observables in Split SUSY have been
presented in~\cite{martin}. The analysis used the $S, T, U$ parameter
expansions, as well as corrections from non-zero momentum
summarized in $Y,V,W$ parameters~\cite{STU}.
They found that the precision electroweak data are
compatible with the Split SUSY spectrum for the values of gaugino and
higgsino masses above the direct collider limits. 
Moreover, Split SUSY corrections to precision observables after LEP2, and by 
considering also the contributions of LEP1 only, are studied in~\cite{strumia}.
For LEP2, the SM prediction fits better than 
Split SUSY predictions, but the difference between the two fits is not
``spectacular''.  
For the LEP1 analysis, on the contrary, the description of the data fits
better in Split SUSY than in the SM (but not dramatically).

The analysis of virtual effects of the 
additional non-standard particles on new physics models
to precision observables requires a very high precision 
of the experimental results as well as of the theoretical predictions.
A predominant role in this respect has to be assigned to the
$\rho$-parameter~\cite{rho}, with loop contributions $\De\rho$
through vector-boson self-energies, which constitute the leading 
process-independent quantum corrections 
to electroweak precision observables, such as the prediction for
$\De r$, the $\MW$--$\MZ$~interdependence,
and the effective leptonic weak mixing angle, $\sweff$. 
Radiative corrections to the electroweak precision observables
within the MSSM have been extensively 
discussed (for a review see, e.g.~\cite{hollik}). 
In particular, a detailed analysis of the SM and the MSSM predictions 
in the $\MW$--$\sweff$ plane, by considering the prospective accuracies 
for the Large Hadron Collider (LHC) and the International Linear Collier
(ILC) with GigaZ option, is included in~\cite{sven}. 
The authors found that the MSSM is slightly favoured over the SM, depending of
the central value of the experimental data.
Once the $W$ mass, the effective leptonic weak mixing angle, $\sweff$,
 as well as the top-quark mass, crucial in this analysis, become known
 with better accuracy at future colliders, a very high precision of the 
theoretical predictions for these observables from both SM and new
physics is needed.

Now we study the effects of gauginos and higgsinos on the 
$\MW$--$\sweff$~interdependence in Split SUSY, i.e. 
when the scalar superpartner masses are too heavy. We focus on the
comparison of  Split SUSY predictions with the SM and MSSM
predictions, by considering the present data and the prospective
experimental precision at the next generation of colliders.
Even if the regions of parameter space allowed by colliders constraints
is expected to be allowed by precision electroweak constraints in Split
SUSY~\cite{martin,strumia}, an analysis of these two observables,  
which are very precisely determined by experiments~\cite{mwsferrors1,mwsferrors}, 
has not yet been done,
and could provide extra
information about the compatibility and/or similarities and differences between
Split SUSY predictions on these two EW precision observables and the SM
and MSSM predictions. 
A few more words are in order with respect the recent analysis in
Refs.\cite{martin,strumia}. The authors of Refs.\cite{martin,strumia}
focus on the 
analysis of current experimental data, performing a $\chi^2$ fit, and
finding whether Split SUSY fits better current experimental
data than the SM. Our work focusses on the possibility of detecting the
deviations induced by Split SUSY in the present and future measurements
of $\MW$ and $\sweff$.

Since the Higgs-boson enters the two electroweak precision
observables we are interested in (by virtue of its contributions to the
self-energies of electroweak vector bosons) an analysis of the radiative
corrections to  the Higgs scalar boson mass from Split SUSY must be
included in our study. It is already known that the strong constraints on
the parameters of low-energy SUSY imposed by the lower bound on the
Higgs-boson mass, $m_H>114.4\GeV$~\cite{mHexp}, 
are relaxed in Split SUSY. This is thanks  to the
large corrections to this mass, due to the renormalization group evolution
from the scale of 
heavy scalars to the weak scale~\cite{split2}. These effects have been
taken into account in our analysis by using the renormalization group
evolution as given in Ref.~\cite{split2}.

Precisely measured observables such as the $W$-boson mass, $\MW$, 
and the effective leptonic mixing 
angle, $\sweff$, are affected by shifts according to
\begin{equation}
\de\MW \approx \frac{\MW}{2}\frac{\cos^2\theta_W}{\cos^2\theta_W - \sin^2\theta_W} \De\rho, \quad
\de\sweff \approx - \frac{\cos^2 \theta_W \sin^2\theta_W}{\cos^2 \theta_W - \sin^2\theta_W} \De\rho\,,
\label{precobs}
\end{equation}
$\theta_W$ being the weak mixing angle, and the electroweak $\rho$
parameter given by
$\De\rho = \frac{\Si_Z(0)}{\MZ^2} - \frac{\Si_W(0)}{\MW^2}$, 
with $\Si_{Z,W}(0)$ the unrenormalized $Z$ and $W$ boson self-energies
at zero momentum.  
We remark that, beyond the $\De\rho$ approximation,
the shifts in these two observables, entering through 
self-energy corrections, are given in terms of the $\de(\De r)$ quantity.
However, the computation and discussion of contributions to $\De r$ in
Split SUSY reduces to the corresponding analysis of the $\De\rho$
quantity. 
In general $\De r$ is given in terms of the photon vacuum polarization,
the ratio of the strengths of neutral and charged currents at vanishing
momentum transfer ($\De\rho$), and the remainder vertex and boxes
contributions. However, if we are interested in extra contributions to
$\De r$ that are not in the SM, each non-SM contribution to the $3$-
and $4$-point functions contains at least one scalar particle. This
scalar can be either the lightest Higgs-boson and, therefore, is like an SM
contribution; or a heavy Higgs-boson or a slepton, whose  
contribution is negligible, since these scalar particles have a mass
of ${\cal O}(10^9\GeV)$ in Split SUSY. As a consequence, no extra contribution
to $\De r$ emerges in this model, and the analysis can be reduced to the 
computation of $\De\rho$ contributions. 

For our computation, we have used \texttt{ZFITTER}~\cite{zfitter} for
the SM prediction. The MSSM contributions to $\De r$ have been taken
from Ref.~\cite{sola}, and we have used
\texttt{Feyn\-Arts}/\texttt{Form\-Calc}/\texttt{Loop\-Tools}~\cite{FAFCLT} for
the vertex contributions to $\sweff$.
The Higgs-boson mass is computed according to Ref.~\cite{split2} for Split
SUSY, and using the leading $\mt, \mb\tb$ approximation for the
MSSM~\cite{Dabels}. The Split SUSY/MSSM contributions to $\De r$ are
added to the \texttt{ZFITTER} computation, and we proceed in an
iterative way to 
compute $\MW,\sin^2\theta_W$. As for the input parameters, we have used
$\MZ=91.1876\GeV$, $\alpha^{-1}(0)=137.0359895$~\cite{pdb},
$\Delta\alpha^{5}_{\rm had}(\MZ)=0.02761\pm0.00036$~\cite{Altarelli}
(corresponding to $\alpha^{-1}(\MZ)=128.936$), 
$\alpha_s(\MZ)=0.119\pm0.003$~\cite{Altarelli}. For the top-quark mass,
we use the latest combination of RunI/II Tevatron data: 
$m_t=172.7\pm 2.9\GeV$~\cite{CDFD0lastmt}. 

The parameter space of Split SUSY is formed by the higgsino mass
parameter $\mu$, the electroweak gaugino soft-SUSY-breaking mass
parameters $M_1$ and $M_2$ (we use the GUT mass relation 
$M_1=M_2 \,5/3\,\tan^2\theta_W$), 
the gluino soft-SUSY-breaking mass $M_g$,
the ratio between the vacuum expectation values of the two Higgs
doublets $\tb=v_2/v_1$, and the scale of the scalar particles masses
$\tilde{m}$. The most important phenomenological consequence of Split
SUSY is the presence of a long-lived
gluino~\cite{split1,split2,collidersplit,gluinolimit}. The scalar mass scale
($\tilde{m}$) lays between the EW scale ($\sim 1\TeV$) and the
unification scale ($\sim 10^{16} \GeV$), current limits from gluino
cosmology set an upper bound $\tilde{m}\lesssim
10^{9}\GeV$~\cite{gluinolimit}. In our computation 
the gluino
mass ($M_g$) and the scalar scale ($\tilde{m}$) enter the Higgs-boson mass
computation, the latter defining the matching scale with the SUSY
theory, and the former through the running of the top quark Yukawa
coupling. For definiteness, we will use $\tilde{m}=10^9 \GeV$, while
$M_g$ is let free.

\begin{figure}[t]
\begin{center}
\begin{tabular}{cc}
\includegraphics*[height=0.42\textwidth]{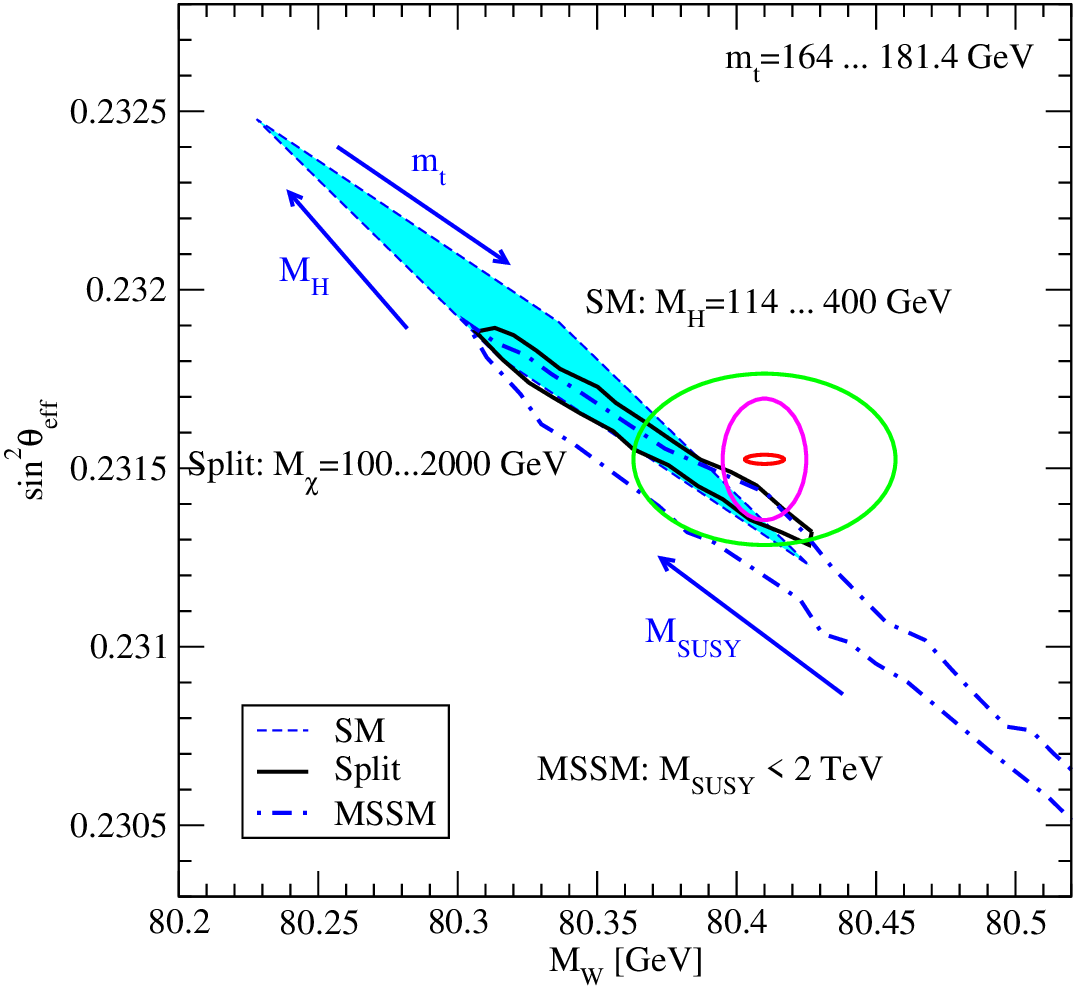}&
\includegraphics*[height=0.42\textwidth]{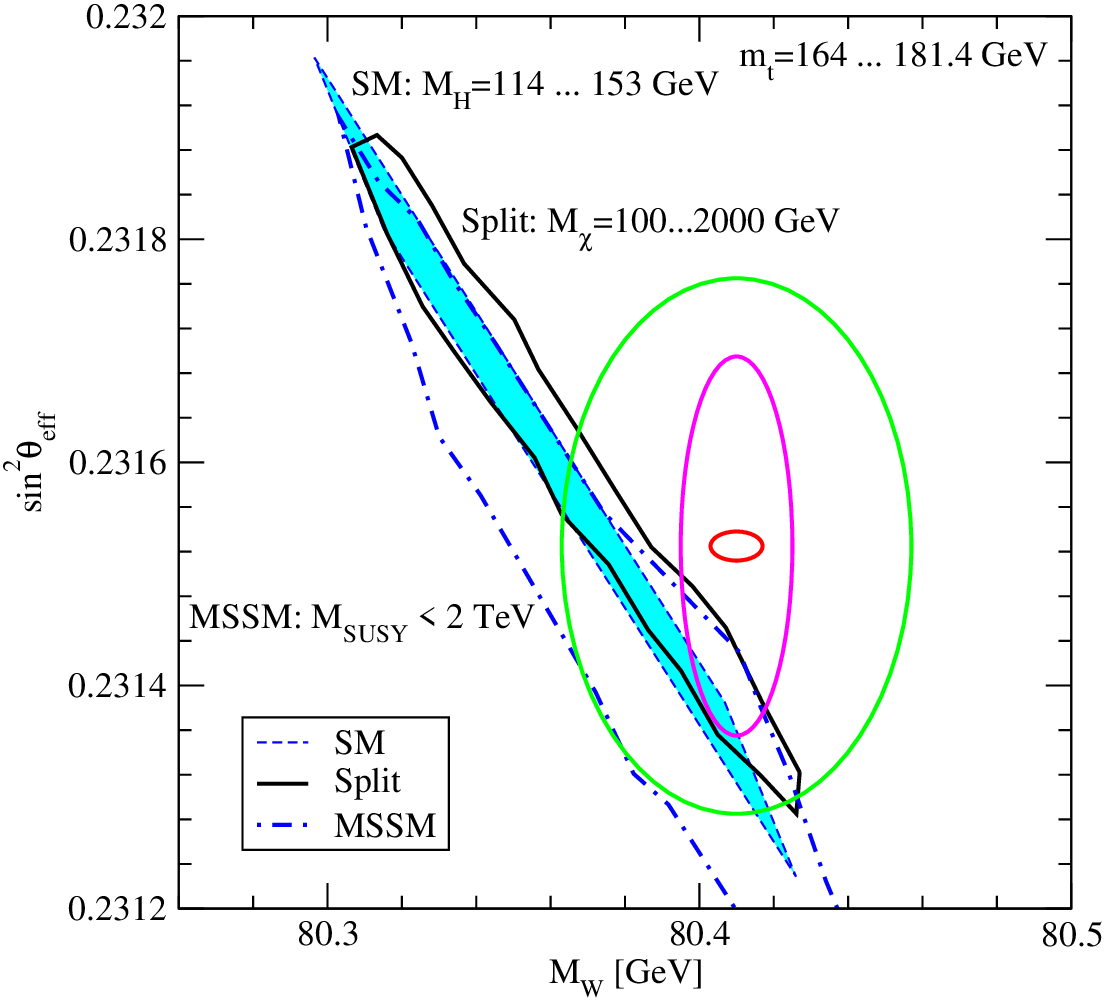}\\
(a) & (b)
\end{tabular}
\end{center}
\caption{SM, MSSM and Split SUSY
predictions for $\MW$ and $\sweff$. The ellipses are 
the experimental results
for $\MW$ and $\sweff$ and the prospective accuracies at 
LEP2/SLD/Tevatron (large ellipse), LHC/ILC (medium ellipse) and
GigaZ (small ellipse).}
\label{fig:SfMwsplit}
\end{figure}

Now we focus on the comparison for $\MW$ and $\sweff$ predictions from
different models with the present data  and the prospective
experimental precision at the next generation of colliders.  
The results for the SM, the MSSM and Split SUSY predictions are given
in~\reffi{fig:SfMwsplit}, in the $\MW$--$\sweff$ plane. The top-quark
mass is varied in the $3\sigma$ range of the current experimental
determination.  
Predictions are shown together with the
experimental results for $\MW$ and $\sweff$ 
(using the current central values: $\MW=80.410\pm 0.032\GeV\,,\,
\sweff=0.231525 \pm 0.00016$) and the prospective
accuracies at present (LEP2/SLD/Tevatron) and at the next generations of
colliders (LHC/ILC and the GigaZ option)~\cite{mwsferrors1,mwsferrors}. 
Our results agree with previous ones for the SM and the MSSM predictions 
given in~\cite{sven}. 

First, we concentrate on results given in~\reffi{fig:SfMwsplit}a. 
We have performed a Monte Carlo scan of the respective parameter space
of the different models, taking into account current experimental limits
on new particles, to find the allowed region in the $\MW$--$\sweff$ plane
for each model. The allowed regions are those enclosed by the
different curves. The arrows show the direction of change in these
regions as the given parameters grow.
The shaded region corresponds to the SM prediction, and it arises from
varying the mass of the SM Higgs-boson, 
from $114\GeV$~\cite{mHexp} to $400\GeV$. 
The region enclosed by the dash-dotted curve corresponds to the
MSSM. Here the SUSY masses are varied between $2 \TeV$ (corresponding to
the upper edge of the area) and close to their experimental 
lower limit $m_{\chi}\gtrsim 100\GeV$, $m_{\tilde f}\gtrsim 150\GeV$
(lower edge of the band).  
As is very well known, contrary to the SM case, 
the lightest MSSM Higgs-boson mass is not a free parameter. 
Thus, the overlap region between SM and MSSM  
corresponds to the region where the Higgs-boson is light, i.e. in the MSSM 
allowed region $m_{h^0} < 140 \GeV$~\cite{hollik},  all
superpartners being heavy (decoupling  
limit in the MSSM), as already established in~\cite{sven}. 
The Split SUSY prediction is summarized in the region enclosed by the
black line in this figure.  
Here, the scalar particles masses are of the order
of $10^9 \GeV$, and the Higgs-boson mass is computed by
following the equations of the renormalization group evolution as
in~\cite{split2}. The computed Higgs-boson mass varies in the range
$m_H^{\rm split}\sim 110$--$153\GeV$. The region excluded by the
experimental Higgs-boson mass limit $m_H\lesssim 114\GeV$~\cite{mHexp}
corresponds 
to a tiny corner of the parameter space: $\tb < 1.5$, $\mt < 166\GeV$.
As expected, we found 
overlap regions between Split SUSY and both the SM and the MSSM. 
Moreover, we see that most of the region predicted
by Split SUSY for $\MW$ and $\sweff$ overlaps with predictions 
already given by the SM and the MSSM.

\begin{figure}[t]
\begin{center}
\includegraphics*[height=0.42\textwidth]{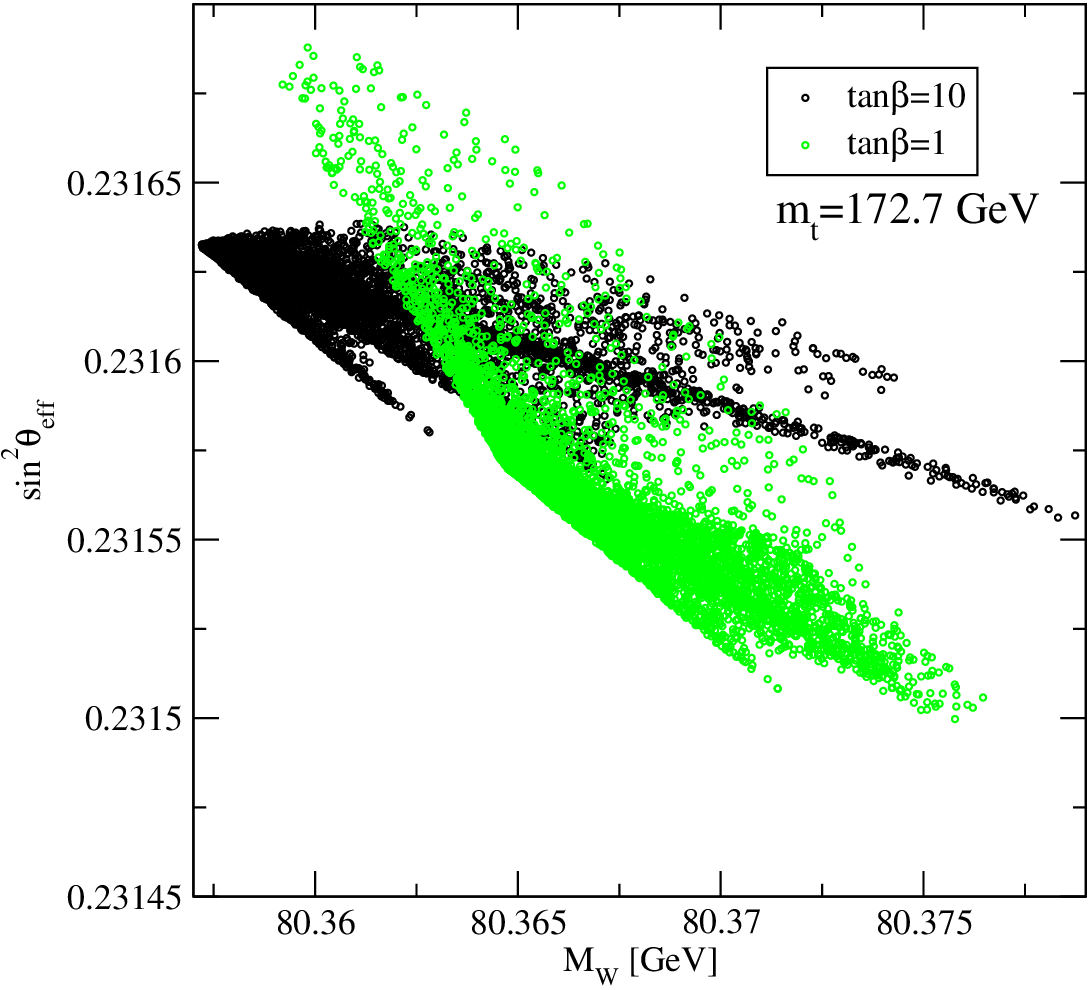}
\end{center}\vspace*{-0.4cm}
\caption{Split SUSY 
predictions for $\MW$ and $\sweff$, with $m_t=172.7\GeV$
and $\tan \beta=1$ (green/light-grey area) and  $\tan \beta=10$
(black area).}
\label{fig:scanmttb}
\end{figure}

In order to clarify the differences of predictions induced by the
three models, we focus on the analysis of the region in which they
overlap. The corresponding results are shown
in~\reffi{fig:SfMwsplit}b (notice the different scales of
the two plots in this figure). 
Here the SM prediction (shaded area)
is fixed to be the one obtained when the SM Higgs-boson mass is varied
in the range of the Split SUSY prediction $m_H=114$--$153 \GeV$. It allows
the extraction of the exact overlap region 
between SM and Split SUSY predictions, by assuming the same Higgs-boson mass
value in the two models. The MSSM results remain as before. 
The region in which the MSSM and Split SUSY overlap 
corresponds to having heavy scalar particles 
(decoupling of squarks, sleptons and extra Higgs bosons in the MSSM) and a
Higgs-boson mass around $140\GeV$ (upper edge of the dash-dotted area). 
However, this small region 
that does not exist in the MSSM emerges in Split SUSY
from the fact that we have light charginos and neutralinos with very
heavy scalars,
 and the Higgs-boson mass is not constrained to be 
$m_{h^0} < 140\GeV$ when all 
superpartners are heavy, as in the MSSM.  
So, there is a new region containing allowed values for the Higgs-boson
mass, $m_H \sim 140$--$153\GeV$, which does not exist in the MSSM. 
On the other hand, the comparison of predictions of Split SUSY with
the SM ones also leads to overlap regions between them. This region  
corresponds to a region with same values of the Higgs-boson mass in the two
models.
Since the region emerging from the Split SUSY predictions is larger than
that obtained from the SM, and by taking into account the experimental
errors, the former might be slightly favoured 
 (depending on the central experimental value). 
Even if we are concerning with just $\MW$ and
$\sweff$ electroweak precision observables, 
this result could also be  extracted from the
analysis of radiative corrections to observables at LEP1 given
in~\cite{strumia}.

\begin{figure}[tp]
\begin{center}
\begin{tabular}{cc}
\includegraphics*[width=0.46\textwidth]{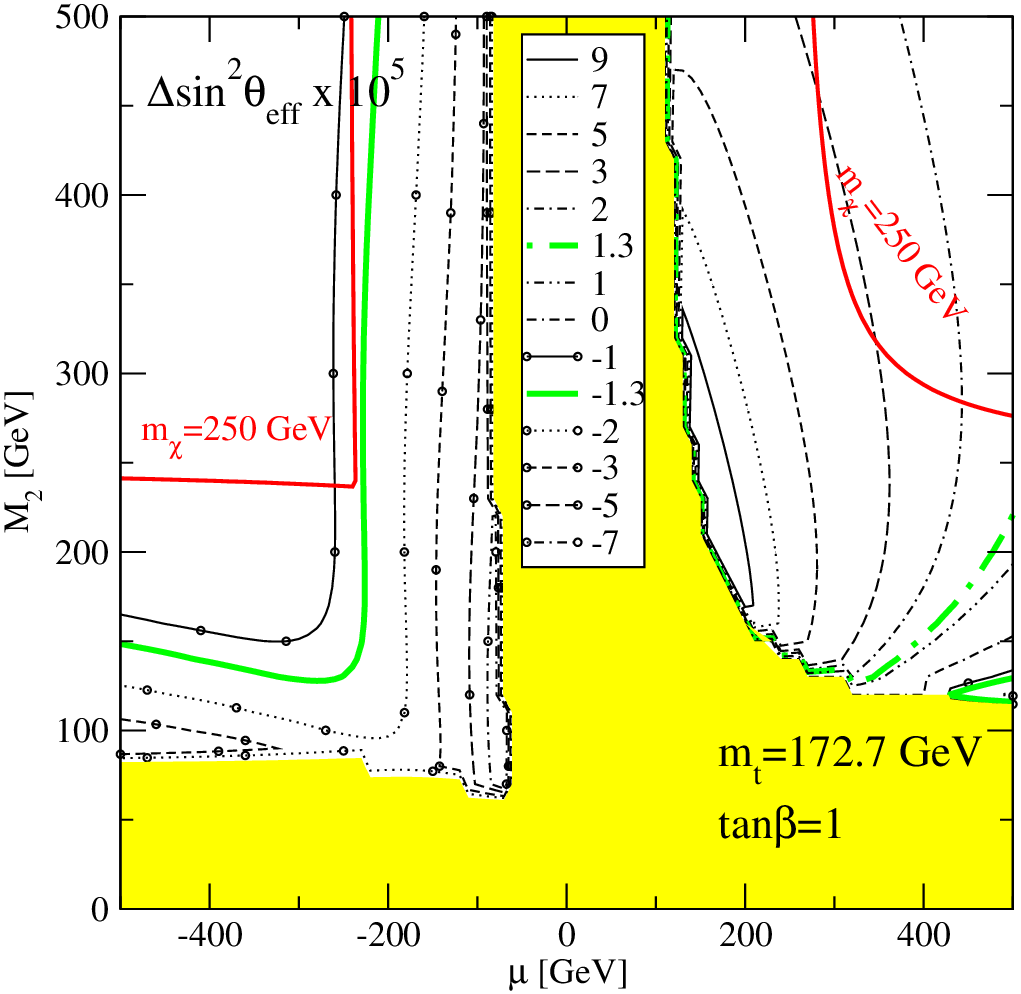} &
\includegraphics*[width=0.46\textwidth]{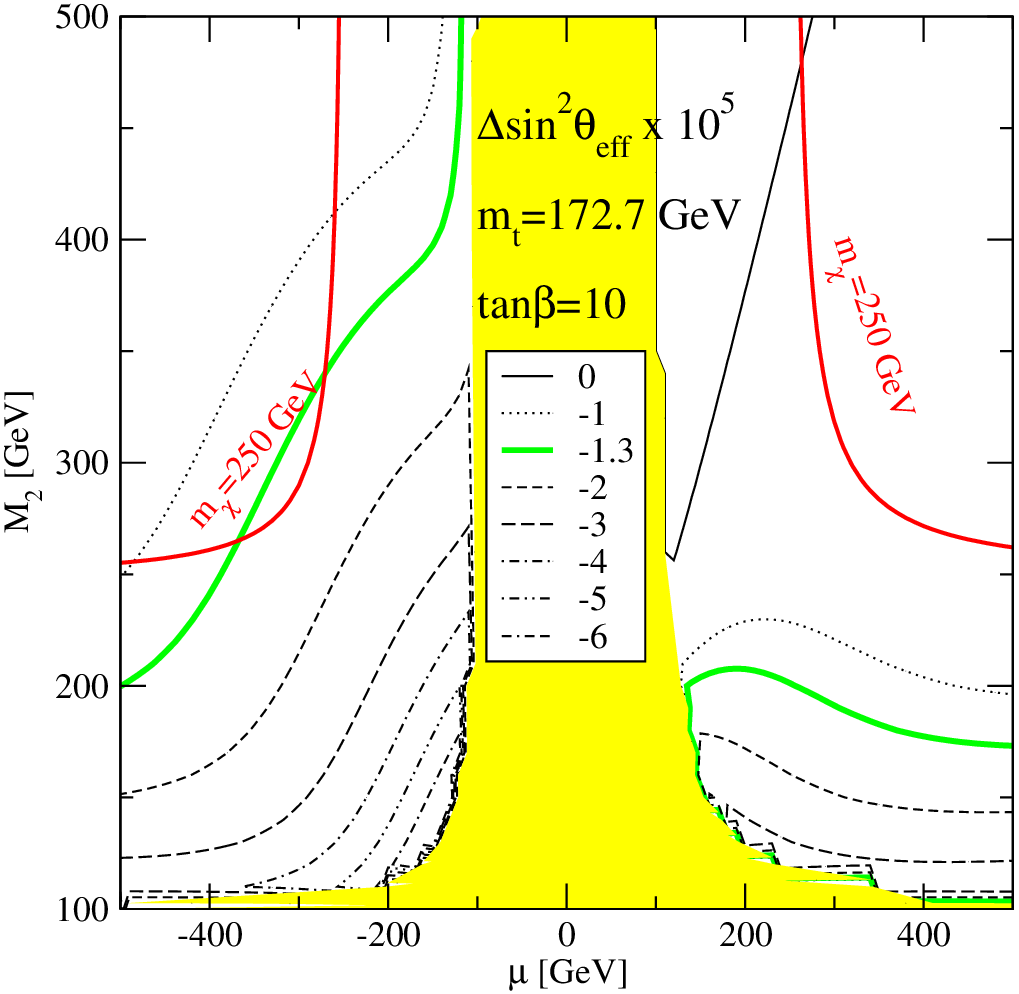}\\
(a) & (b) \\
\includegraphics*[width=0.46\textwidth]{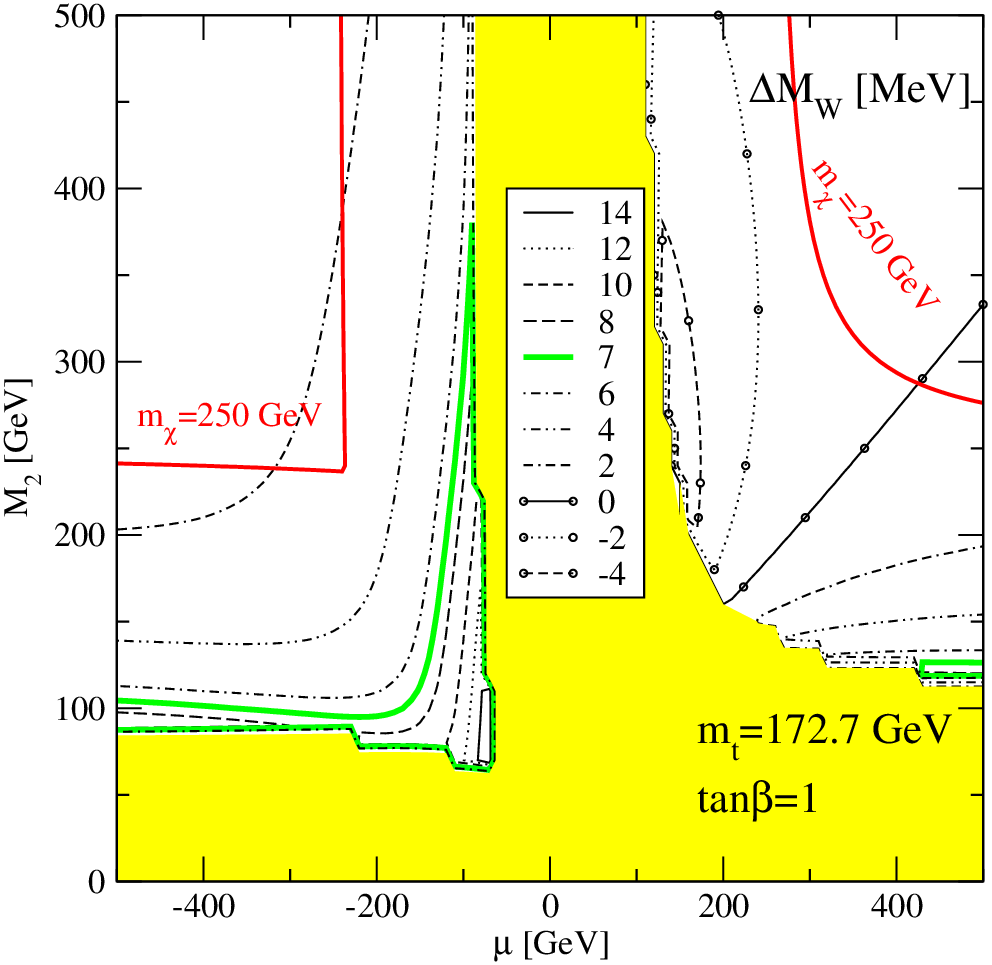} &
\includegraphics*[width=0.46\textwidth]{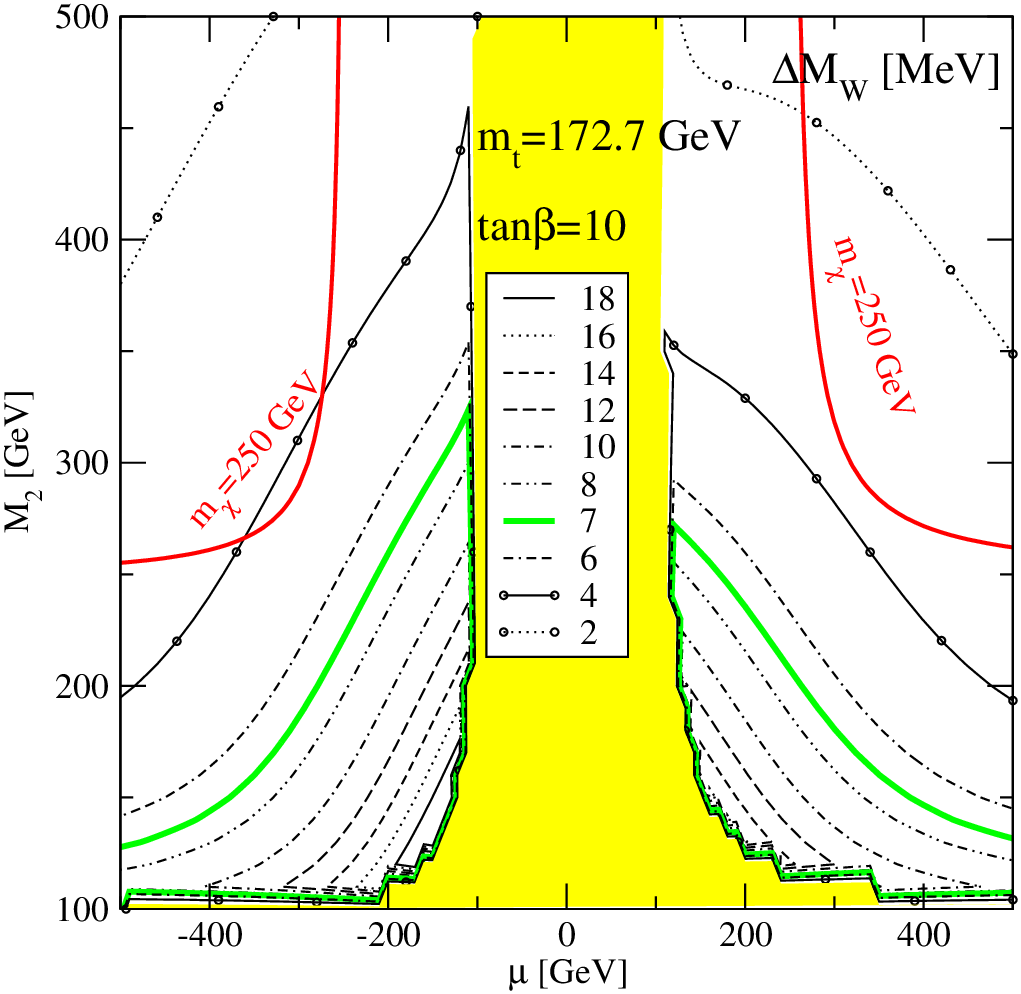} \\
(c) & (d) 
\end{tabular}
\end{center}
\caption{The shifts $\De \sweff$ and $\De \MW$ in the $[M_2$--$\mu]$ 
plane for $m_t=172.7\GeV$
  and for $\tan \beta=1$ (a, c) and $\tan \beta=10$ (b, d). 
  The shaded region corresponds to $m_\chi<100\GeV$. Also shown is the
  line corresponding to a lightest chargino mass $m_\chi=250\GeV$. 
  The gluino mass is taken to be $M_g=500\GeV$.}
\label{fig:mudepen}
\end{figure}

From now on, we focus on the 
differences between SM and Split SUSY predictions. To assess the
importance of the Split SUSY contributions, we must compare these 
with the present and future experimental uncertainties and SM
theoretical errors. The current experimental
uncertainties are~\cite{ewdataw03}
\begin{equation}
\Delta \MW^{\rm exp,today} \approx 34 \mev, \qquad
 \Delta \sweff^{\rm exp,today} \approx 17 \times 10^{-5} \,;
\label{eq:exptoday}
\end{equation}
the expected experimental precision for the LHC is~\cite{MWatLHC}
\begin{equation}
\Delta \MW^{\rm LHC} \approx 15\mbox{--}20 \mev\,\,;
\label{eq:expLHC}
\end{equation}
and at a future linear
collider running on the $Z$ 
peak and the $WW$ threshold (GigaZ) one expects~\cite{mwsferrors1,moenig}
\begin{equation}
\Delta\MW^{\rm exp,future} \approx 7 \mev,  \qquad
 \Delta \sweff^{\rm exp,future} \approx 1.3 \times 10^{-5} ~.
\label{eq:expfuture}
\end{equation}
On the other hand, the theoretical intrinsic uncertainties in the SM
computation are~\cite{hollik}:
\begin{eqnarray}
\Delta\MW^{\rm th, today,SM} \approx 4 \mev,  &\qquad&
 \Delta \sweff^{\rm th,today,SM} \approx 5 \times 10^{-5}~,\nonumber\\
\Delta\MW^{\rm th, future,SM} \approx 2 \mev,  &\qquad&
 \Delta \sweff^{\rm th,future,SM} \approx 2 \times 10^{-5} ~.
  \label{eq:theorerror}  
\end{eqnarray}

We remark that the radiative corrections induced by Split SUSY in the 
$\MW$ and $\sweff$ precision observables and, in particular, their
differences with respect to predictions from other models, 
depends strongly on the Higgs-boson mass. 
Besides, the
role of the $\tan \beta$ parameter in this analysis is dictated by the fact
that the Higgs-boson mass increases with $\tan \beta$ for small values
of this parameter, around $1$--$5$. For larger values of $\tan \beta$, 
we found that the Higgs-boson mass remains stable. 
Figure~\ref{fig:scanmttb} shows the result of the parameter scan in Split
SUSY for the central experimental value of the top-quark mass
$m_t=172.7\GeV$, and two different values of $\tb$.
The results obtained when taking $\tan \beta=1$ are displayed in the
green/light-grey  area  
of this figure. The black area represents $\tan \beta=10$.
The Higgs-boson mass runs between
$114\GeV$ and $153\GeV$, as predicted by Split SUSY. 
We can see that the effective leptonic 
weak mixing angle, $\sweff$, always decreases when $\tan \beta=10$ but,
on the contrary, its value increases  when $\tan \beta=1$ for some specific set of 
values of the other parameters, in particular when $\mu>0$ (see below). 
So, the correction
to $\sweff$ is positive for small values of $\tan \beta$ and $\mu>0$. 
The corrections to $\MW$ are positive over a large range of the
parameter space. When $\tan \beta=1$  and $\mu>0$ we can also get
negative corrections. We found that for values of $\tan \beta$ larger
than $10$, the above conclusions remain unchanged. 

In~\reffi{fig:mudepen}
we show the shifts $\De \sweff$ and $\De \MW$ 
in the $[M_2$--$\mu]$ plane.
The shifts in the variables are defined as: 
$\De X\equiv X^{\rm Split\ SUSY} - X^{\rm SM}$, where the SM computation
is performed using the Higgs-boson mass predicted by Split SUSY.
The top-quark mass is fixed to its central value $m_t=172.7\GeV$, while 
$\tan \beta=1$ in Figs.~\ref{fig:mudepen}a,c and $\tb=10$ in
Figs.~\ref{fig:mudepen}b,d.  
The regions with a chargino mass smaller than $100\GeV$ are excluded.  
At the upper side of this figure we display the shifts on the
effective leptonic weak mixing angle, $\De \sweff$ and, in the lower
side the results for $\De \MW$. 
The Split-SUSY-induced shifts are $|\De \sweff|<10 \times 10^{-5}$ and
$|\De\MW|<20\MeV$; as of today's data~(\ref{eq:exptoday}) they are
smaller than the experimental error, and the data cannot discriminate
between the SM and Split SUSY. The same conclusion applies to the
accuracy reached at the LHC~(\ref{eq:expLHC}). However, the shifts are
larger than the experimental accuracy of GigaZ~(\ref{eq:expfuture}) in
certain regions of the parameter space. 
For $\tb=1$, the shift in $|\De \sweff|$ is
larger than $1.3\times 10^{-5}$ for most of the explored region for
$\mu>0$ and for the region with $\mu<0$: $\mu\gtrsim-250\GeV$ or
$M_2\lesssim 150\GeV$ (Fig.~\ref{fig:mudepen}a). At $\tb=10$
(Fig.~\ref{fig:mudepen}b), $|\De \sweff|$ is larger than the future
experimental accuracy~(\ref{eq:expfuture}) in
a small region $M_2\lesssim 175$--$200\GeV$ for $\mu>0$, and a large region
$M_2\lesssim 200$--$500\GeV$ for $\mu<0$. 
As far as $\MW$ is concerned, the LHC measurement~(\ref{eq:expLHC})
could only be useful in a small corner of the parameter space for
$\mu<0$, $\tb\gsim10$. The GigaZ measurement~(\ref{eq:expfuture}) does
not help for $\tb=1$, $\mu>0$, owing to the cancellation of the corrections
in the center of the region. For $\tb=1$, $\mu<0$ there exists a small region
for $M_2\lesssim 110\GeV$ or $\mu>-110\GeV$. For larger $\tb$, the region
of sensitivity is much larger. Summarizing the results of
Fig.~\ref{fig:mudepen}:
\begin{itemize}
\item Positive shifts of $\sweff$ are only possible at
small $\tb\simeq1$ and $\mu>0$.  They are large, and  correlated
with small and negative shifts of $\MW$. These large shifts are possible
even for large values of the chargino masses ($m_\chi > 250\GeV$).
\item For $\tb\simeq1$, $\mu<0$ large negative shifts in $\sweff$ are
  possible, correlated with positive shifts in $\MW$, but $\sweff$ is
  the most sensitive of those observables.
\item For large $\tb\gtrsim 10$ and $\mu>0$, the sensitivity region is
  confined to small $M_2\lesssim 275$--$375\GeV$, with the largest shift
  provided by $\sweff$ for $\mu\gtrsim300\GeV$, and by $\MW$ otherwise. 
\item Finally, for large $\tb\gtrsim10$ and $\mu<0$, the largest
  sensitivity is provided by $\sweff$; it can reach GigaZ sensitivities
  even for moderate chargino masses ($m_\chi\approx250\GeV$).
\end{itemize}
We would like to stress the fact that the results for negative $\mu$
are quite different from those of positive $\mu$. As
Fig.~\ref{fig:mudepen} shows, changing the sign of $\mu$ can change
the sign and the absolute value of the shifts significantly, so
conclusions derived from an analysis of the $\mu>0$ scenario only do not
necessarily apply to the complete Split SUSY parameter space.

\begin{figure}[t]
\begin{center}
\includegraphics*[height=0.45\textwidth]{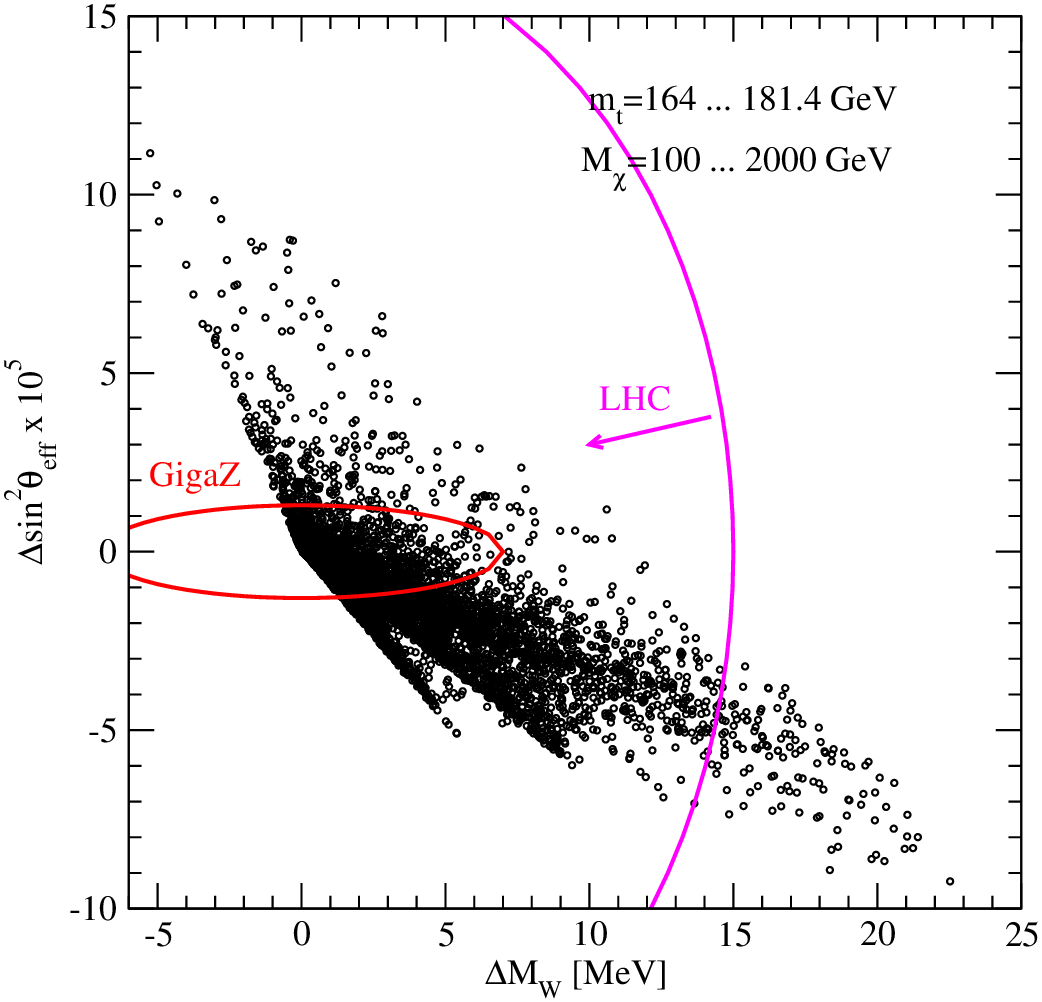}
\end{center}
\caption{Shifts of the differences between Split SUSY and SM 
predictions for $\MW$ and $\sweff$, scanning over the parameter space. 
Also shown are  the ellipses for the  prospective accuracies at 
LHC/ILC (large ellipse) and
GigaZ (small ellipse).}
\label{fig:shifts}
\end{figure}

To finish the discussion on the shifts $\De \sweff$ and $\De \MW$, 
the results of the difference between
Split SUSY and SM predictions in the  $\MW$--$\sweff$ plane
are displayed in~\reffi{fig:shifts}, together with the expected error ellipses
of the future colliders~(\ref{eq:expLHC}) and (\ref{eq:expfuture}) centered at the SM value. 
These variations for $\sweff$ and $\MW$
have to be compared with the numbers of
eqs.~(\ref{eq:exptoday})--(\ref{eq:theorerror}). 
We can see that the shift $\De \MW$ can be up to $23\MeV$ at
its maximum and, therefore, it is impossible to discriminate between models
at the present experimental accuracy. However, future
experiments could be probed with the future precision on $\MW$, 
if theoretical uncertainties will be sufficiently
under control. 
On the other hand, the shifts $\De \sweff$ can easily reach values  
$\pm 2 \times 10^{-5}$, which is larger than both the expected experimental
errors and the anticipated theoretical accuracies~(\ref{eq:theorerror}).

As a side note, we observe from Fig.~\ref{fig:SfMwsplit}b that the
current SM prediction of $\MW$--$\sweff$ would need a positive shift on
both observables (together with a large value of $\mt$) to be closer to
the central experimental value. Figs.~\ref{fig:mudepen},
\ref{fig:shifts} show that the general trend of the Split SUSY
contributions is a negative correlation of the shifts
on both observables, that is, if $\De\MW>0$ then $\De\sweff<0$, an
reciprocally. The region providing ($\De\MW>0$, $\De\sweff>0$) is
actually small and the largest region
corresponds to ($\De\MW>0$, $\De\sweff<0$)
--c.f. Fig.~\ref{fig:shifts}. Of course, since we are dealing with high
precision observables, small deviations from the general trend are
important, and Refs.~\cite{martin,strumia} actually show that there are
points of the parameter Split SUSY space that fit better than the SM
the experimental value of the electroweak precision observables.

In conclusion, we have computed the Split SUSY contributions to the
electroweak 
precision observables $\MW$ and $\sweff$ arising from a heavy scalar
spectrum and light charginos and neutralinos. 
For the computation, we have evaluated the Higgs-boson mass by using
the renormalization-group evolution equations, the
$\Delta r$-parameter, and then $\MW$ and $\sweff$. 
Numerically we compared the effects of radiative corrections to these
observables induced by Split SUSY, SM and the MSSM, and with present and
future experimental and theoretical accuracies. We find that the shifts
induced in Split SUSY models are smaller than
present experimental accuracies~(\ref{eq:exptoday}), and therefore no
conclusion can be drawn with respect to the validity of this model. With
the anticipated LHC accuracy on $\MW$, a small corner of the parameter
space can be explored. However, only with the GigaZ option of the
ILC the experiment would be sensitive to the Split SUSY
corrections to these observables. In this option, the effective leptonic
mixing angle ($\sweff$) is the most sensitive of the two observables. For
moderate and large $\tb$, the lightest chargino must be relatively light,
$m_\chi \lesssim 250\GeV$, which should have already been detected
either at the LHC or the ILC before the GigaZ era. The observables provide, however,
a high-precision test of the model. An interesting case is 
a scenario with low $\tb\simeq1$ and positive $\mu$, where large shifts
in $\sweff$ are expected, even for large values of the chargino masses.


\subsection*{Acknowledgements}
\vspace{-.5em}
We thank N.~Arkani-Hamed, G.~F.~Giudice and A.~Romanino for very
helpful discussions. 
The work of S.P. has been supported by the European Union under contract
No.~MEIF-CT-2003-500030, and that of 
J.G. by a \textit{Ramon y Cajal} contract from MEC (Spain) and in part by
MEC and FEDER under project 2004-04582-C02-01.
J.G. is thankful for the warm hospitality at CERN, and S.P. is thankful
for the warm hospitality at the Universitat de Barcelona, where part of
this work was done.


\end{document}